\documentclass[osajnl,twocolumn,showpacs,superscriptaddress,10pt]{revtex4-1} 
\usepackage{amsmath,amssymb,graphicx}
\begin{document}

\title{Polarization-resolved sensing with tilted fiber Bragg gratings: theory and limits of detection}

\author{Aliaksandr Bialiayeu}\email{Corresponding author: alexbeliaev@gmail.com}
\affiliation{Department of Electronics, Carleton University, 1125 Colonel By Drive, Ottawa, ON K1S5B6, Canada}
\author{Anatoli Ianoul}
\affiliation{Department of Chemistry, Carleton University, 1125 Colonel By Drive, Ottawa, ON K1S5B6, Canada}
\author{Jacques Albert}
\affiliation{Department of Electronics, Carleton University, 1125 Colonel By Drive, Ottawa, ON K1S5B6, Canada}

\begin{abstract}
Polarization based sensing with tilted fiber Bragg grating (TFBG) sensors is analysed theoretically by two alternative approaches. The first method is based on tracking the grating transmission for two orthogonal states of linear polarized light that are extracted from the measured Jones matrix or Stokes vectors of the TFBG transmission spectra. The second method is based on the measurements along the system principle axes and polarization dependent loss (PDL) parameter, also calculated from measured data. It is shown that the frequent crossing of the Jones matrix eigenvalues as a function of wavelength leads to a non-physical interchange of the calculated principal axes; a method to remove this unwanted mathematical artefact and to restore the order of the system eigenvalues and the corresponding principal axes is provided.
A comparison of the two approaches reveals that the PDL method provides a smaller standard deviation and therefore lower limit of detection in refractometric sensing. Furthermore, the polarization analysis of the measured spectra allows for the identification of the principal states of polarization of the sensor system and consequentially for the calculation of the transmission spectrum for any incident polarization state. The stability of the orientation of the system principal axes is also investigated as a function of wavelength. 
\end{abstract}



\maketitle 

\section{Introduction}
Polarization-based sensing is crucial for various types of optical sensors, in particular for stress analysis, plasmon-mediated sensing, and sensing of anisotropic media or other forms of perturbations~\cite{Zhang:12, Cranch:06, Kotov:13, Wang:13}. In the particular case of waveguide-type sensors (including optical fibers), it is possible to interrogate the device optical properties with polarized light but in general this require very careful alignment and control of the input polarization (which is relatively easy in waveguides but not so much in conventional fibers). Here we describe a data analysis method that provides a full characterization of the polarization-resolved optical transmission state of any sensor system based on measurements of the wavelength dependence of the Jones matrix with an Optical Vector Analyser (OVA). The OVA measures the transmission amplitude and phase (or reflection) of an optical system as a function of polarization and wavelength and provides the Jones matrix elements for each wavelength. We use a standard tilted fiber Bragg grating (TFBG) refractometric sensor as a test system for demonstrating the new approach. The results further confirm that the polarization dependent loss parameter (PDL) of the system provides a smaller standard deviation (and hence better limit of detection) than the transmission spectra extracted along the principal axes, as initially observed empirically in \cite{Caucheteur:11}.

\begin{figure}[htbp]
\centerline{\includegraphics[width=.8\columnwidth]{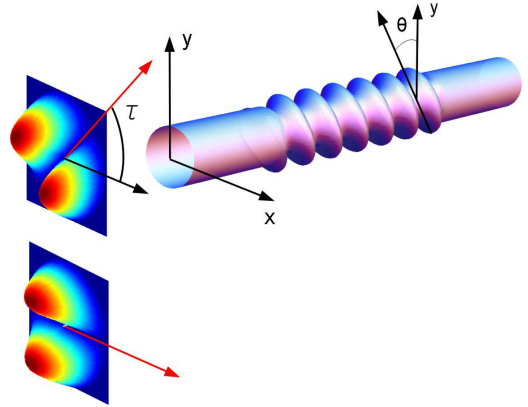}}
\caption{The schematic representation of TFBG grating and the incident linearly polarized core mode (the $E_\rho$ component is shown). The core mode is rotated about the optical device axis by some angle~$\phi$. 
The grating is tilted by $\theta$ angle about the $\hat{x}$ axis.}
\label{Fig0}
\end{figure}

It is known that TFBGs have strong polarization-dependent properties~\cite{Caucheteur:2008, Bialiayeu:2012, Alam:13} due to the tilt of the grating planes which breaks the cylindrical symmetry of the fiber and strongly impacts the magnitude of the coupling coefficients between the incident \textbf{$HE_{11}$} core mode and the cladding modes of various polarizations (\textbf{TE, TM, HE, EH} modes) excited by the grating~\cite{Lee_00}, as is schematically shown in Fig.~\ref{Fig0}. 

\begin{figure}[htbp]
\centerline{\includegraphics[width=.95\columnwidth]{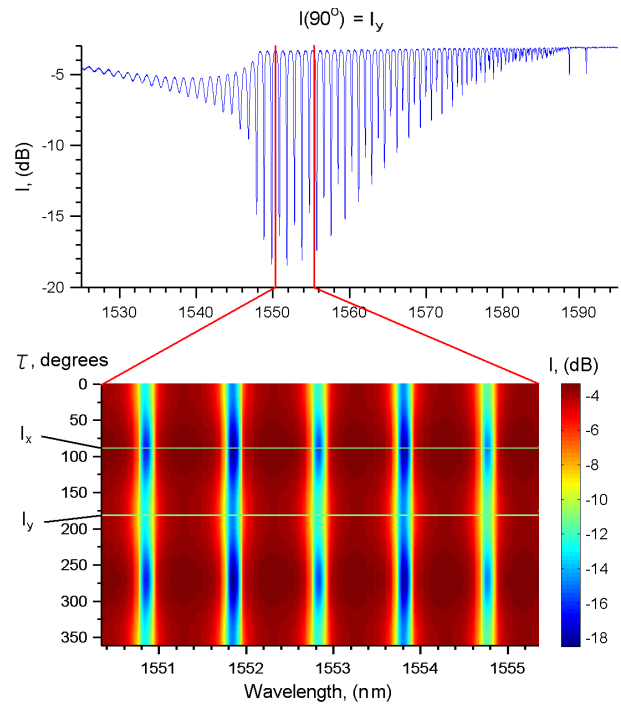}}
\caption{A typical TFBG transmission spectrum for linearly polarized light, and series of spectra obtained by rotating a linear polarizer about the optical axis are shown as the density plot.}
\label{Fig1}
\end{figure}

In particular, for a grating tilted by an angle $\theta$ in the reference $y-z$~plane and a linearly polarized input core mode whose polarization is rotated by an arbitrary angle~$\phi$ with the reference $x$ axis (a most general occurrence in optical fibers that are not polarization preserving), the coupling to individual cladding modes will depend strongly on both $\theta$ and $\phi$. It has been further demonstrated recently that high order cladding modes of the \textbf{EH} and \textbf{TM} families have radially polarized evanescent fields around the fiber cladding while \textbf{HE/TE} modes are azimuthally polarized. This is important because in many sensing applications (or other forms of interactions between cladding guided light and coatings or external media, such as nonlinear switching~\cite{Villanueva:11}, or four wave mixing~\cite{Shao:2012} applications), the state of polarization of the evanescent field can be a determinant factor. This most easily seen for metallic coatings whose boundaries conditions depend strongly on the polarization of the incident electric field relative to the plane of the boundary~\cite{Caucheteur:11, Bialiayeu:2012}. It was also demonstrated in \cite{Alam:13} that only radially polarized cladding modes are excited by TFBGs when the input core mode is linearly polarized along the tilt plane (\textit{i.e.} along $y$), while the orthogonal input polarization (along $x$) generates only azimuthally polarized cladding modes. For convenience and for easy comparison with conventional optics, in the remainder of this paper we use "\textbf{TE}" to refer to the \textbf{HE} and \textbf{TE} fiber mode families (azimuthally polarized,~\textit{i.e.} polarized along the plane of the fiber cladding boundary) and "\textbf{TM}" for the \textbf{EH} and \textbf{TM} fiber mode families (radially polarized,~\textit{i.e.} perpendicularly to the fiber cladding surface). This is further justified by the fact that on the scale of the wavelength ($1.5~\mu m$, the curvature of the cladding boundary (diameter of $125~\mu m$) is negligible and the boundary can be viewed as approximately flat surface. 
The effect of the input mode polarization can be observed by measuring optical transmission spectra with a polarizer inserted between the light source and the grating~\cite{Bialiayeu:2011}. A typical series of spectra measured at various angles of linearly polarized incident light is shown in Fig.~\ref{Fig1} for a $1~cm$ long TFBG immersed in water.

It is clear from Fig.~\ref{Fig1} that there are two orthogonal states of transmission, oriented at $90^o$ degrees relative to each other, where the spectrum has well-defined extrema. We can therefore find the polarizer angles for which~\textbf{TE} and~\textbf{TM} modes are excited~\textit{a posteriori} without prior knowledge of the actual orientation of the tilt plane. However this is often impractical, especially if the device under test has to be moved between measurements, thereby potentially altering the light polarization orientation between the polarizer and the TFBG. A much more practical and rapid method to interrogate such devices is to use an OVA to get the Jones matrix elements at each wavelength and to use the information to calculate the full set of polarization properties of the system, including the transmission along the chosen geometrical axes ($I_x$ and $I_y$ in Fig.~\ref{Fig1}) as well as the transmission along the system principal axes and PDL spectrum. How this is done is presented in the next section.

\section{Measurements along principal axes of an optical system}
In this section we review the PDL technique and provide an approach to study the system transmission spectrum along its principal axes. 
 
A linear optical system can be represented in terms of the ${2 \times 2}$ Jones matrix $[J]$ connecting the incident and transmitted electric field vectors~\cite{JONES:41}: 
\begin{equation}
 \vec{E}_{out} = [J] \vec{E}_{in}
\end{equation}

In practice, the transmission spectrum of a device under test is usually characterized by two major parameters: a polarization-independent parameter called the insertion loss ${I(\lambda)}$ and the polarization-dependent loss parameter ${PDL(\lambda)}$, both measured as a function of the free space wavelength $\lambda$.
To extract these parameters from the Jones matrix, a hermitian matrix $[H]$ has to be constructed first.
\begin{equation}
[H] = [J]^\dagger [J]
\end{equation}
Next, carrying out the eigen decomposition of the $[H]$ matrix
\begin{equation}
[H] = [U] [\Lambda] [U]^T
\end{equation}
allows us to find a diagonal matrix $[\Lambda]$ containing the eigenvalues $\rho_1$ and $\rho_2$ of the $[H]$ matrix.

Alternatively, the singular value decomposition (SVD) of the Jones matrix can be computed~\cite{Heffner:1992}:
\begin{equation}
[J] = [U] [\Sigma][V]^T
\end{equation}
where the diagonal matrix $[\Sigma]$ contains two singular values $\sigma_1$ and $\sigma_2$ such that ${\rho_1 = \sigma_1^2}$ and ${\rho_2 = \sigma_2^2}$, due to the fact that ${[H] = [J]^\dagger [J]}$.

The matrix ${[U] = (\vec{u}_1, \vec{u_2})}$ resulting from the SVD contains two orthogonal vectors corresponding to $\rho_1$ and $\rho_2$ eigenvalues, and these eigenvectors correspond to the device response when the electric field vector $\vec{E}$ is aligned with the system principal axes,~\textit{i.e.} $\vec{E} \parallel \vec{u}_k$ of the system. The eigenvalues $\rho_k =\sigma_k^2$ can be interpreted as the observable transmission loss for the corresponding incident polarizations.
The principal axes $\vec{u}_1$, $\vec{u_2}$ of an arbitrary fiber device are showed schematically in Fig.~\ref{Fig3}. The angle $\phi$ corresponds to the rotation between the arbitrarily selected frame of reference $(x,y)$ and the principal axes defined by $\vec{u}_1$, $\vec{u_2}$.
\begin{figure}[htbp]
\centerline{\includegraphics[width=.65\columnwidth]{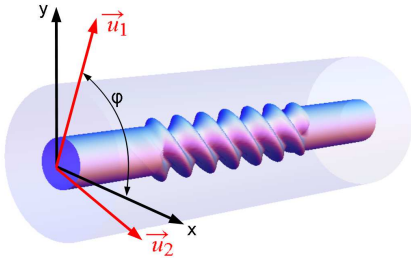}}
\caption{TFBG with physical $\hat{x}$, $\hat{y}$ axes and principal system axes $\vec{u}_1$, $\vec{u}_2$ measured at some wavelength $\lambda$.}
\label{Fig3}
\end{figure}

Having the singular values of the Jones matrix $[J]$ or eigenvalues of ${[H] = [J]^\dagger [J]}$ matrix allow for the computation of the insertion loss, which is the average loss over all possible polarizations, as follows~\cite{Heffner:1992}.
\begin{eqnarray}
\label{Iout}
I_{out} & =& \langle \vec{E}_{out}|\vec{E}_{out} \rangle = \langle \vec{E}_{in}|[J]^\dagger [J]| \vec{E}_{in} \rangle = \nonumber \\
& =& \langle \vec{E'}_{in}|\Lambda | \vec{E'}_{in} \rangle = \frac{\rho_1 + \rho_2}{2} I_{in}
\end{eqnarray}
Which can be re-formulated in the more frequently used dB scale as:
\begin{equation}
\frac{I_{out}}{I_{in}} = 10 \log_{10}{\frac{\rho_1 + \rho_2}{2}},
\end{equation}
here, the scalar product is denoted in accordance with Dirac's bra–ket notation. 
Similarly, the polarization dependent loss, which is the magnitude of the difference between the maximum and minimum device transmission over all possible input polarization actually corresponds to the difference in the loss measured along the system principal axes $\vec{u}_1$ and $\vec{u}_2$. It is defined as follows:
\begin{equation}
PDL = 10 |\log_{10}{\frac{\rho_1}{\rho_2}|.}
\end{equation}

Alternatively we can introduce a polarization parameter with a linear scale \cite{Bialiayeu:2012}:
\begin{equation}
P = \frac{|\rho_1 - \rho_2|}{\rho_1 + \rho_2}.
\end{equation}
 
\begin{figure*}[htbp]
\centerline{\includegraphics[width=1.7\columnwidth]{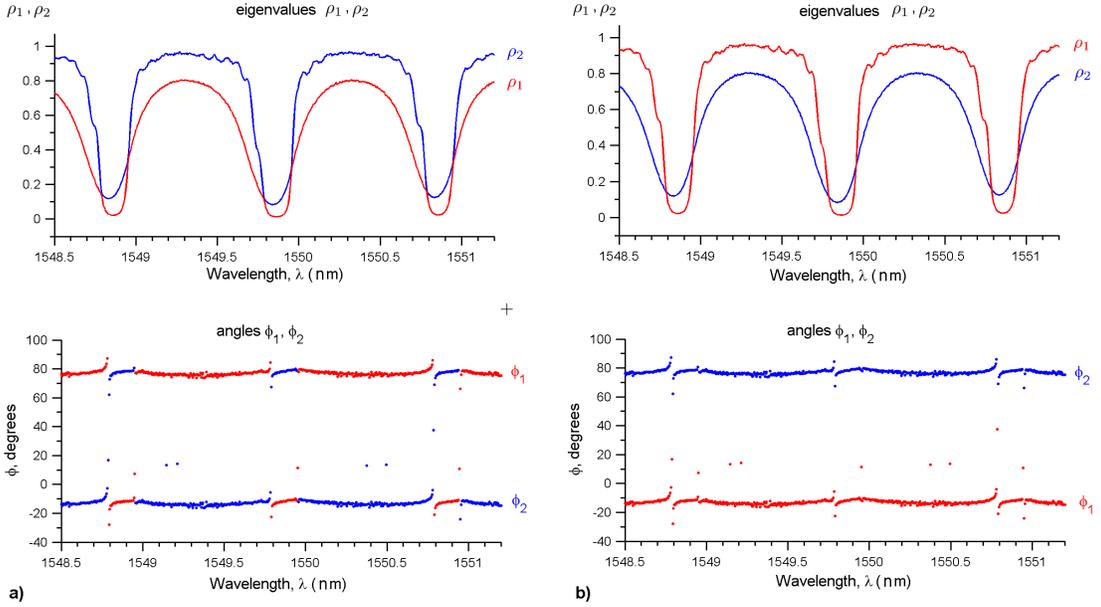}}
\caption{Eigenvalues $\rho_1$ and $\rho_2$ and the angle $\phi$ between the geometrical axes of the system $\hat x$, $\hat y$ and the coordinate system defined by the principal axes $\vec{u}_1$, $\vec{u}_2$, before a) and after b) the eigenvalues were reordered. (The data were obtained by means of the OVA 5000 Luna Technologies.)}
\label{Fig3}
\end{figure*}

In addition to the eigenvalues of the system transmission matrix (plotted as a function of wavelength in Fig.~\ref{Fig3}), we can also find the angle of rotation~($\phi$) of the principal axes about the device optical axis~($z$) and plot it as a function of wavelength (bottom frame of Fig.~\ref{Fig3}). 
From the mathematical point of view, both eigenvalues are roots of quadric polynomial and can be order arbitrary, usually in descending order: this is why $\rho_1$ is always larger than $\rho_2$ in Fig.~\ref{Fig3}a, hence the eigenstates are interchanged when they cross. Because of this effect, the corresponding eigenvectors are interchanged as well, therefore the angle $\phi$ experiences ${\pi/2}$ shifts at the crossing points, as shown on the bottom panel of Fig.~\ref{Fig3}a. Therefore, a strategy for restoring the individual transmission spectra along the principal axes is to interchange the eigenvalues and principal axes every time the ${\phi = \pi/2}$ jumps are detected. 

The result of such reordering is shown Fig.~\ref{Fig3}b. Now the transmission spectrum corresponds to linearly polarized light with its electric field vector aligned with each principal axis of the system and the rotation angle jumps are eliminated.

\begin{figure}[htbp]
\centerline{\includegraphics[width=.95\columnwidth]{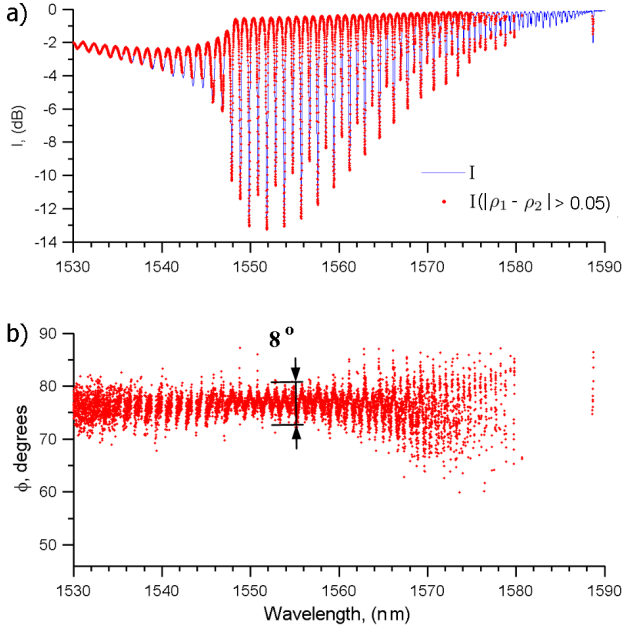}}
\caption{Rotation of the principal axes of TFBG device as a function of optical wavelength. (The data was obtained by means of the OVA 5000 Luna Technologies.)}
\label{Fig4}
\end{figure}

In general, the principal axes of an optical system are not necessarily fixed with respect to a reference frame but may depend on the wavelength, as shown in Fig.~\ref{Fig4} for a TFBG sensor. 
Indeed, by changing the optical frequency it is expected that an optical system would operate differently. The global behavior of the principal axes, along the whole operational range of the TFBG sensor, is shown in Fig.\ref{Fig4}b, and the corresponding insertion loss in Fig.\ref{Fig4}a.
To remove the noise only the points at which the difference between eigenvalues is noticeable (${|\rho_2 - \rho_1| > 0.05}$) are plotted. The noise arises from the fact, that at the points of eigenvalues crossing the transmission matrix become degenerate, has the eigenvectors are not well defined, and oscillate rapidly with respect to the geometrical axis, as it can be seen from Fig.~\ref{Fig4}b at the points of crossing.

Figure~\ref{Fig4}b shows that the system possesses significant polarization asymmetry, or birefringence, in the wavelength range ${\lambda \in [1545-1575]}$. Although the principal axes are globally stable, near $75^o$ degrees relative to the reference frame of the LUNA interrogation system, locally they experience small oscillations, of about $8^o$ degrees about the optical axis. These oscillations are related to the resonances observed in the spectrum (Fig.~\ref{Fig4}a), and reflect a wavelength dependent birefringence.

We also note the observed alternation between the peaks (Fig.~\ref{Fig4}b), which can be explained by the fact that the alternating peaks have different azimuthal symmetries and polarizations~\cite{Lee_00}.


\section{Extracting transmission spectra for light polarized in an arbitrary direction from the Jones matrix or the Stokes vector data}
In this section we describe an approach allowing to extract the transmission loss spectra for light that is linearly polarized in any direction, which may be a useful predictive tool for device performance under real life conditions without the need for an external polarizer.

\subsection{The Jones matrix analysis}

The Jones matrix of an optical device under test connected to a linear polarizer is given by:
\begin{equation}
[J_{dev + pol}] = [J_{dev}][J_{pol}].
\end{equation}
Here ${[J_{dev}]}$ is Jones matrix of a device measured by an optical vector analyzer and $[J_{pol}]$ is the Jones matrix of a linear polarizer, oriented at an arbitrary angle theta relative to the system frame of reference given by:
\begin{eqnarray}
[J_{pol}(\theta)] &=& \begin{bmatrix} \cos^2(\theta) & \cos(\theta)\sin(\theta) \\ \sin(\theta)\cos(\theta) & \sin^2(\theta) \end{bmatrix},
\end{eqnarray}

Thus, on optical device with a known Jones matrix ${[J_{dev}]}$, connected to a linear polarizer rotated by $\theta$ angle, is described by the following Jones matrix:
\begin{equation}
[J_{dev + pol}(\theta)] = [J_{dev}] \begin{bmatrix} \cos^2(\theta) & \cos(\theta)\sin(\theta) \\ \sin(\theta)\cos(\theta) & \sin^2(\theta) \end{bmatrix}.
\end{equation}

And finally the average transmission spectrum ${I(\theta,\lambda)}$, for a given angle~$\theta$ of the linear polarizer, can be computed as 
\begin{equation}
I(\theta,\lambda) = 10 \log_{10}{\frac{\rho_1(\theta,\lambda) + \rho_2(\theta,\lambda)}{2}},
\end{equation}
where ${\rho_1(\theta,\lambda)}$ and ${\rho_2(\theta,\lambda)}$ are the eigenvalues of ${H(\theta,\lambda) = [J_{dev + pol}(\theta,\lambda)]^\dagger [J_{dev + pol}(\theta,\lambda)]}$ matrix.

\subsection{The Stokes vector analysis}
As a supplementary information, we provide here an alternative approach to calculating the response of a device to arbitrarily oriented linearly polarized light using the Stokes vector of the system. The Stokes vector can be measured with a polarization controller or with a dedicated instrument, such as the JDS Uniphase SWS-OMNI-2 system.

A beam of light can be completely described by four parameters~\cite{Berry:77, Goldstein:1992}, represented in the form of the Stokes vector:
\begin{equation}
\vec{S} = \begin{pmatrix} S_0 \\ S_1\\ S_2\\ S_3 \end{pmatrix} = \begin{pmatrix} I(0^o)+I(90^o) \\ I(0^o)-I(90^o) \\ I(45^o)-I(135^o) \\ I_{RHS} - I_{LHS} \end{pmatrix},
\end{equation}
here $I(\theta)$ is the measured power transmission coefficient of the light polarized in the direction defined by the angle $\theta$ in the plane perpendicular to the direction of light propagation, and $I_{RHS}$, $I_{LHC}$ are the coefficients for right- and left-handed circular polarized light, respectively.

Since the Stokes parameters are dependent upon the choice of axes, they can be transformed into a different coordinate system with a rotation matrix $[R]$.

\begin{equation}
[R(\theta)] = \begin{bmatrix} \cos(\theta) & -\sin(\theta) \\ \sin(\theta) & \cos(\theta) \end{bmatrix},
\end{equation}

Considering that a second coordinate system is obtained by rotating the original coordinate system about the direction of light propagation on the angle $\theta$, we can write~\cite{McMaster:1954} 

\begin{equation}
\vec{S'}(\theta) = [R(\theta)] \vec{S},
\end{equation}
or
\begin{eqnarray}
\begin{pmatrix} S_0' \\ S_1'\\ S_2'\\ S_3' \end{pmatrix} &=& \begin{bmatrix} 1 & 0 & 0 & 0 \\ 0 & \cos(2\theta) & \sin(2\theta) & 0 \\ 0& -\sin(2\theta) & \cos(2\theta) & 0 \\ 0 & 0 & 0& 1 \end{bmatrix} \begin{pmatrix} S_0 \\ S_1 \\ S_2 \\ S_3 \end{pmatrix} = \nonumber \\
 &=& \begin{pmatrix} S_0 \\ S_1 \cos(2\theta) + S_2 \sin(2\theta) \\ -S_1 \sin(2\theta) + S_2 \cos(2\theta) \\ S_3 \end{pmatrix},
\end{eqnarray}

Now considering that
\begin{eqnarray}
I(0^o)+I(90^o) &=& S_o, \nonumber \\
I(0^o)-I(90^o) &=& S_1 \cos(2 \theta) + S_2 \sin(2 \theta),
\end{eqnarray}
we conclude that the transmission loss spectra along the two orthogonal axes, rotated by the angle $\theta$ with respect to the original system of measurements, are given by the following expressions:
\begin{eqnarray}
I_x(\theta,\lambda) = \frac{1}{2}(S_o(\lambda) + S_1(\lambda) \cos(2 \theta) + S_2(\lambda) \sin(2 \theta)), \nonumber \\
I_y(\theta,\lambda) = \frac{1}{2}(S_o(\lambda) - S_1(\lambda) \cos(2 \theta) - S_2(\lambda) \sin(2 \theta)). \nonumber \\
\end{eqnarray}
Hence, if the Stokes vector is know in one coordinate system, the transmission spectrum of linearly polarized light with the electric field $\vec{E}$ aligned along an arbitrary angle $\theta$, can be recovered.

\section{Comparison between measurements along the geometrical and principal axes}



\begin{figure}[htbp]
\centerline{\includegraphics[width=.75\columnwidth]{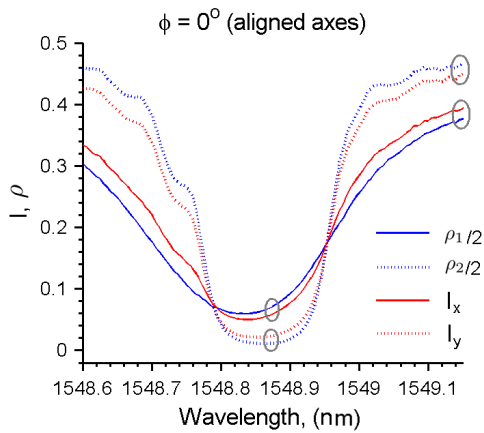}}
\caption{Transmission losses along TFBG system principal axes and its geometrical axes for perfectly aligned coordinate systems ${\phi = 0^o}$.} 
\label{Fig5}
\end{figure}

The last results show that the transmission spectra of light linearly polarized along any direction in the transverse plane of the optical axis can be computed from the Jones matrix or from the Stokes vector, in addition the possibility of measuring it directly by rotating a polarizer, as shown in Fig.~\ref{Fig1}. 
In particular, the transmission spectra of light linearly polarized along the system principal axes can be extracted from the Jones matrix and compared to the direct measurement with a polarizer. 
Figure~\ref{Fig5} shows a comparison between the direct measurement of one resonance, obtained by aligning the polarizer along the average direction of the principal axes (for a given spectrum), and the eigenstate spectra obtained from the Jones matrix. 
The small difference in the two results comes from the aforementioned wavelength dependent oscillation of the principal axes (which cannot be compensated for in the direct measurement), and to a possible slight change in polarization state between polarizer and the grating (since a non-polarization maintaining fiber is used). The following section will demonstrate that in spite of this small inaccuracy, the parameters extracted from the Jones matrix data provide excellent spectral sensitivity results for refractometric sensing.

\section{Polarization-based detection of small refractive index changes with TFBG sensors}
In this section we investigate which polarization-based measurement techniques provide the best signal-to-noise ratio when TFBG sensors are used to detect small refractive index changes, and use the special case of a TFBG coated with gold nanorods (as described in~\cite{Bialiayeu:2012}). This choice is made because the polarization dependence of waveguide-type sensors is much enhanced when metal coatings are used. As indicated earlier, the interaction of guided waves with metal interfaces depends strongly on whether the electric fields of the waves are tangential or normal to the metal boundary. It was further mentioned that when the core-guided input light of a TFBG is linearly polarized along the principal axes, the electric fields of high order cladding modes are either tangential or radial (hence normal) to the cladding boundary. To be precise, $y$-polarized input light (corresponding to light polarized parallel to the plane of incidence on the tilted grating fringes, as shown in Fig.~\ref{Fig0},~\textit{i.e.} P-polarized) couples to radially polarized cladding modes while $x$-polarized light (perpendicular to the tilt plane, or S-polarized) couples to azimuthally polarized cladding modes (tangential to the boundary). Further polarization effects arise with non-uniform metal coatings, since they have boundaries that are both tangential and radially oriented relative to the cylindrical geometry of the fiber~\cite{Zhou:13}. It is therefore desirable to carry out two transmission measurements along the principal axis to observe directly such polarization effects on the strengths and positions of the cladding mode resonances~\cite{Berini:2011}. On the other hand, it has been shown here that a Jones matrix measurement can provide this information as well, in addition to other parameters of interest, such as PDL. While the PDL spectrum "hides" the physical effects responsible for difference in transmission due to the different polarization states, it has been shown in the past to yield excellent limits of detection for surface plasmon resonance based TFBG sensors~\cite{Caucheteur:11}. We now proceed to compare the signal noise for refractive index measurements by the various polarization dependent data extraction techniques.

\begin{figure}[htbp]
\centerline{\includegraphics[width=.8\columnwidth]{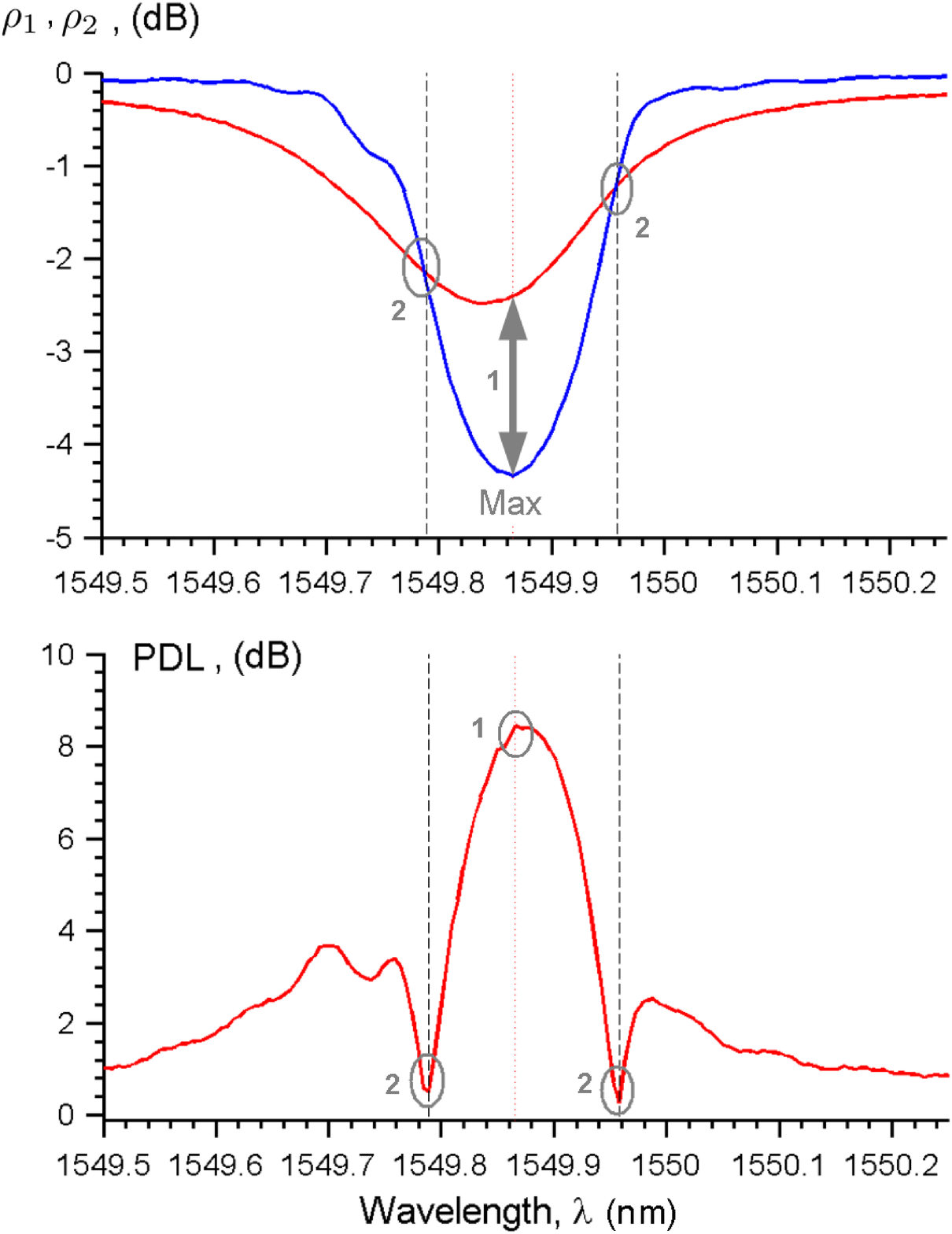}}
\caption{Two eigenvalue spectra (individual transmission along the principal axes) and corresponding polarization dependent loss (PDL) parameter. The peak in PDL spectrum is denoted by "1" and zeros by "2". Here the more common dB scale is used.}
\label{Fig6}
\end{figure}

\begin{figure}[htbp]
\centerline{\includegraphics[width=1.0\columnwidth]{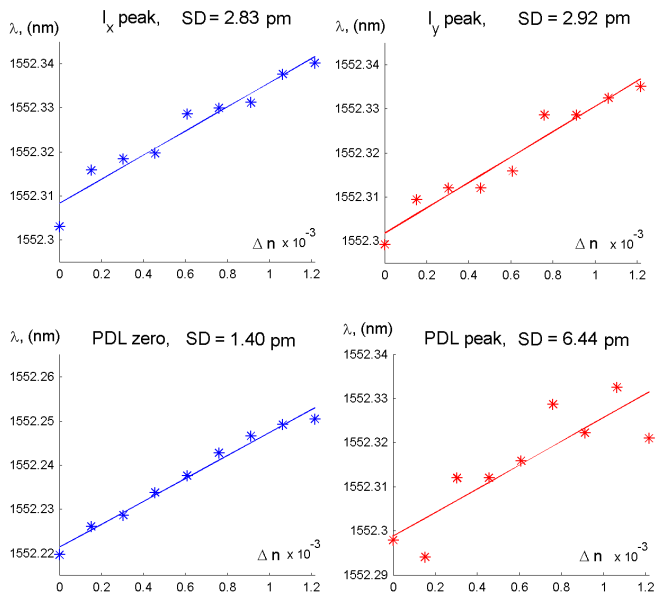}}
\caption{Position of the resonance a) in the transmission spectra $I_x$ and $I_y$ of light polarized along ${\hat x}$ and ${\hat y}$ geometrical axes here aligned with the principal axes, and b) in PDL spectra (the maximum and zero values of PDL are detected), as a function of refractive index change. 
The continuous lines represent the least square approximation to the measured data, and $SD$ is the standard deviation from the linear approximation}
\label{Fig7}
\end{figure}

As shown in Fig.~\ref{Fig6} for a TFBG coated with a sparse layer of gold nanorods and immersed in water, the PDL parameter provides the absolute value of the difference between resonances observed in the transmission spectra measured along the principal axes. 
The relative position of the peaks, their amplitude, and their width become convoluted in the PDL parameter, which provides instead a maximum located somewhat in between the individual resonance maxima and a zero on either side corresponding to the wavelengths where the spectra cross each other. 
When the refractive index of the medium surrounding such TFBG changes the waveguiding characteristics of the cladding are modified and the resonances observed in the transmission spectrum change accordingly. Therefore, to detect changes in refractive index we can either follow the amplitudes and positions of individual resonances~\cite{Berini:2011} or of the PDL features. Here, the sensor was immersed in water and the refractive index was incrementally increased in steps of ${\Delta n =1.517 \times 10^{-4}}$ by adding ${10~\mu l}$ of ethylene glycol ( ${C_2 H_4 (OH)_2}$ ) to ${5~m l}$ to the water. The impact of each increase in refractive index on all parameters of interest is shown in~Fig.\ref{Fig7} for a typical slice of the spectrum (it was noted in \cite{Bialiayeu:2012} that there was little difference in wavelength shifts across the TFBG spectrum for this device).

The refractive index change was chosen to have a relatively small value of (${\Delta n =1.517 \times 10^{-4}}$) to test the sensor detection limits and a linear fitting was used because the sensor response for such small changes is expected to be linear. The standard deviation of the errors from the linear fit was calculated in the usual manner by: 

\begin{equation}
SD = \sqrt{\frac{1}{N} \sum_{i=1}^N (x_i - \mu)^2},
\end{equation}
here $\mu$ is the expected value of $\vec{x}$, and ${\vec{x} = \vec{y}_{appr} - \vec{y}_{data}}$ is the difference between the measured data~$\vec{y}_{data}$ and its linear approximation~$\vec{y}_{appr}$.

The results of the fits show that the most accurate detection of the refractive index change is achieved with the zeros in the PDL spectrum, as this measurement provides the smallest standard deviation of~${SD = 1.40~pm}$. 
The worst result were obtained with the PDL peak (${SD = 6.44~pm}$), while the detection of individual polarized resonances provides an intermediate value of the standard deviation (essentially equal to $3$~pm). It is not surprising that zeros of PDL should provide the most accurate results as they consist essentially of a differential measurement between two spectra with well-defined crossing points that occur on the sides of the individual resonances, where the spectral slope is highest. On the other hand, at the expense of an increase in noise by a factor of approximately 2, other effects such as differences in the change in the resonance amplitude for the two polarizations (which can be linked to differential loss or scattering) can be studied when the principal axis spectra are used.

\section{Conclusion}

In this paper we investigated the use of Jones matrix and Stokes vector based techniques and polarization analysis to extract information from optical fiber sensors in non-polarization maintaining fibers. In particular we showed how to calculate transmission spectra with electric fields $\vec{E}$ aligned with the system principal axes or along any other system axes. 
In the case of a TFBG inscribed in a non-polarization maintaining fiber for instance, the principal axes of the system are determined by the direction of the tilt of the grating planes (and its perpendicular). 
The transmission spectra for light polarized in the tilt plane (P-polarized) or out of the tilt plane (S-polarized) can thus be extracted without having to separately align a linear polarizer upstream from the grating and hoping that the polarization remains linear between the polarizer and the TFBG. Polarization-resolved spectra can also be obtained at a faster rate, for applications in chemical deposition process monitoring for instance, without the need to line up or rotate a polarizer between each measurement. 

The transmission spectra of P- and S-polarized light were compared with the TFBG spectral response along the principal axes. A small oscillation of about 8 degrees in the orientation of the principal axes as a function of wavelength was observed.

Furthermore, it was determined that the best TFBG sensor detection limits are achieved when zeros in the PDL spectrum are followed, mainly because of the sharpness of crossing points occurring on the steep sides of the individual resonances, but at the expense of the direct observation of the sensor response to polarized light (which are also available from the Jones matrix data). In the latter case, while the standard deviation of the spectral sensitivity is doubled relative to PDL measurements, additional information becomes available regarding the influence of the measured medium on cladding mode loss, allowing further uses from the TFBG data sets~\cite{Wenjun:2014}.

\end{document}